# Shot noise detection in hBN-based tunnel junctions

Panpan Zhou,[1] Will J. Hardy,[2] Kenji Watanabe,[3] Takashi Taniguchi,[3] and Douglas Natelson[1,4,5]

[1]Department of Physics and Astronomy, Rice University, 6100 Main Street, Houston, TX 77005, USA

[2]Applied Physics Program, Smalley-Curl Institute, Rice University, 6100 Main Street, Houston, TX 77005, USA

[3]National Institute for Materials Science, 1-1 Namiki, Tsukuba, Ibaraki 305-0044, Japan

[4]Department of Electrical and Computer Engineering, Rice University, 6100 Main Street, Houston, TX 77005, USA

[5]Department of Materials Science and Nanoengineering, Rice University, 6100 Main Street, Houston, TX 77005, USA

**ABSTRACT**

High quality Au/hBN/Au tunnel devices are fabricated using transferred atomically thin hexagonal boron nitride as the tunneling barrier. All tunnel junctions show tunneling resistance on the order of several k$\Omega$/$\mu$m$^2$. Ohmic *I-V* curves at small bias with no signs of resonances indicate the sparsity of defects. Tunneling current shot noise is measured in these devices, and the excess shot noise shows consistency with theoretical expectations. These results show that atomically thin hBN is an excellent tunnel barrier, especially for the study of shot noise properties, and this can enable the study of tunneling density of states and shot noise spectroscopy in more complex systems.

Shot noise refers to the electrical current fluctuations in conductors driven out of equilibrium that originate from the discrete nature of charge carriers. The full shot noise intensity $S$ in normal metal-insulator-normal metal (N-I-N) tunnel junctions has the well-defined result $S = 2eI$ when $eV \gg 2k_BT$, where $I$ is the average current, a result that follows from the Poisson statistics of uncorrelated tunneling of particles of charge $-e$. In more complex systems, where the magnitude of the effective quasiparticle charge $e^*$ is different from $e$, the shot noise would change accordingly[1–4]. Thus, the so-called Fano factor $F = S/2eI$ would deviate from 1.

To achieve high quality tunnel junctions, the insulating layer material is required to be thin and uniform, with low disorder. Traditional metal oxides or high-k dielectric materials that are often used as tunnel barriers can suffer from conducting impurities, defects, pinholes or non-uniform thickness when approaching the subnanometer scale. Hexagonal boron nitride (hBN), an insulating 2D material with a large bandgap (in the range 5.2 ~5.9 eV) and atomically smooth surface, has been shown to be an excellent candidate for a next-generation tunnel barrier. The thickness of the hBN layer is extremely uniform, as is the van der Waals distance between hBN and surrounding materials. When interfaces with adjacent materials are set by the van der Waals interaction, issues of interfacial strain and lattice mismatch for the transferred hBN may not be as critical as in grown oxide tunnel barriers. Compared with high-k dielectric materials (typically transition metal oxides), hBN is very chemically inert, which makes it compatible for integration with a variety of complicated materials. Vertically stacked tunnel devices have been applied to study many interesting phenomena such as field-effect tunneling transistor[5,6], resonant tunneling[7,8], and spin-dependent tunneling[9]. However, shot noise, which is a key aspect of the tunneling process and can identify other transport mechanisms, has not been well studied in the hBN-based tunnel junctions. For ideal shot noise measurements, tunnel junctions with low defect rates and high tunneling current density are required, which is still challenging.

   Here we demonstrate shot noise detection in high quality Au/hBN/Au tunnel junctions from room temperature down to the cryogenic regime. The Au/hBN/Au tunnel junction is fabricated using a wet transfer method that has been widely used in 2D materials[10,11]. When working with monolayer hBN flakes, the wet transfer method here resulted in a much higher yield of devices than the nominally cleaner dry transfer method[12]. To begin, hBN flakes were exfoliated from bulk single crystals onto 300 nm $SiO_2$/Si wafers. With the help of a differential integral contrast (DIC) microscope (Olympus BX60M), atomically thin hBN flakes could be identified by their slight optical contrast with the wafer substrate (Fig. 1(a)). For transfer to another substrate, PMMA 950 (4% in chloroform) was spin coated at 3000 rpm for 1 min on the substrate containing hBN flakes, and the chip was baked at 180℃ for 3 min on a hotplate. Next, a flexible adhesive tape (NITTO SPV-224) was prepared by cutting out a small window about 2×2 mm and positioned with the window over the PMMA-coated thin hBN flake of interest, thus providing a mechanical support for the thin PMMA film. Then, the whole substrate was soaked in 1 M KOH solution for 2 hours at 50℃ to etch the $SiO_2$ layer, releasing the hBN attached to the PMMA membrane from the substrate, followed by cleaning in de-ionized (DI) water for 5 hours. Then, using a micromanipulator stage and microscope, the hBN flake was aligned to another substrate prepared with Au electrodes (Fig. 1(b)). To remove the residual PMMA, the chip was rinsed in acetone and isopropanol and annealed in forming gas (25% $H_2$, 75% $N_2$) at 250℃ for 2.5 h. Finally, top electrodes over the hBN were defined via e-beam lithography followed by evaporation of Ti (2 nm)/Au (30 nm), with the final device as shown in Fig. 1(c). All junctions are around 1×1 μm in size.

The tunneling current I and differential conductance dI/dV were measured simultaneously by applying a voltage excitation $V_{dc}+V_{ac}$, where $V_{dc}$ is a variable DC bias voltage and $V_{ac}$ is 1 $mV_p$ AC at ~500 Hz. For all the monolayer hBN tunnel junctions, the tunnel resistivity ranges from 1-7 kΩ/μm$^2$ and the resistivity almost has no temperature dependence, which agrees with previous studies on hBN based tunneling devices[13,14].

Surface roughness and grain size of both the underlying and overlying Au layers can result in reduction of the "true" tunnel junction area below the lithographically defined area. We believe that this is why the device tunneling resistances per area show small variations from one device to another. The I-V curves are well described as Ohmic over our bias range (-0.1 V to 0.1 V), as shown in Fig. 2(a). The smooth linear curve suggest the low defect densities present in these tunnel junctions, as defect-mediated transport is known to produce Coulomb staircase-like conduction, including strong suppression of tunnel current at small bias and step-like features in the I-V curve[14,15]. Small deviations from Ohmic response can be revealed by measuring G = dI/dV as a function of DC bias, as shown in Fig. 2(b). The differential conductance has small changes (<10%) within the measurement bias range, and there are no clear features corresponding to possible hBN phonon modes for any of the devices. In the case of phonon assisted tunneling processes, in previous experiments involving graphene/hBN tunneling devices kink features have been observed in the dI/dV data, leading to pronounced peaks in the second derivative of the tunnel current, $d^2I/dV^2$, at voltages where known hBN phonon modes exist , as expected for inelastic electron tunneling spectroscopy (IETS). In our experiment, the comparatively high measurement temperature can broaden and wash out any IETS signatures.  Further, remaining chemical residue from the fabrication process may contribute to the observed small variations of the dI/dV curves obscuring any hBN phonon modes.

To measure the shot noise signal in these Au/hBN/Au junctions, a modulated radio frequency (RF) measurement technique[18–20] was adapted and applied here. The schematic of the measurement circuit is shown in Fig. 3(a). A function generator (Stanford Research DS-345) is used to apply a square wave bias switching between zero voltage and a finite voltage at ~5 kHz across the junction sample. In the zero-voltage state, the sample is in equilibrium and there is only thermal noise (Johnson-Nyquist noise) in the system; whereas in the finite-voltage state, the tunnel current will contribute to extra noise. The excess noise is the noise difference between the finite-bias state and the zero-bias state. A bias-tee separates the low frequency and RF current signals collected from the junction. The low frequency ("DC" compared to the rf signal) bias across the device is modulated as a square wave at 5 kHz and supplied via the low frequency port of the bias tee.  The resulting square wave current is collected from the low frequency port of the bias tee on the other side of the device, and measured using a current preamplifier (Stanford Research Systems SR570) and lock-in detection at the modulation frequency. To avoid 1/f noise effects in the system, an RF frequency range that is much higher than the 1/f noise rolloff (tens of kHz) was chosen for the noise detection. The RF component, which contains the fluctuation information, is filtered using a 250 MHz to 600 MHz band-pass filter and then amplified and measured by a logarithmic power detector to convert into a voltage output (which corresponds to the noise power). A second lock-in amplifier also synchronized to the square wave detects the difference between the power detector's output corresponding to finite-voltage and zero-voltage. This difference combined with the detector's average output can finally be translated to give the difference in noise power at finite-voltage and zero-voltage, which is proportional to the excess noise $S_I(V)-S_I(V=0)$. The detected power is ultimately limited by the reflection coefficient $\Gamma = (R - Z_0)/(R + Z_0)$. In the mismatched case where the sample's resistance R is much larger than impedance of the transmission line $Z_0 = 50\Omega$, the noise power $P_{measure}$ that is coupled to the transmission line and amplifier chain becomes[18]

$$P_{measure} = P_{noise}(1 - \Gamma^2) \approx 4Z_0 S_I$$

where $S_I$ is the current noise spectrum density. Ideally, by calculating the measured power, the sample's real current fluctuation can be identified. However, as the effective electrical circuit shows in Fig. 3(b), many additional, extrinsic factors can contribute to RF signal loss, including capacitive coupling to ground, effective inductance of the narrow Au leads and their antenna effects, non-ideal wirebonds and connectors, and loss in coaxial wiring, so the measured signal is smaller than the ideal value by a factor particular to the measurement setup and geometry that must be calibrated. Moreover, care in interpretation must be taken in devices with strongly nonlinear I-V response, as the analysis above assumes linearity. As implemented, this modulated RF shot noise measurement technique can be applied to samples with weakly non-linear I-V curves, and the sensitivity is limited by the measurement system's RF signal attenuation and the power detector's background fluctuation.

The evolution of shot noise with temperature has the well-known form[18,19]:
$$S_I = 2eVG \coth\left(\frac{eV}{2k_BT}\right)$$

In the low temperature limit where $\frac{eV}{2k_BT} \gg 1$, $S_I \approx 2eVG = 2eI$, which gives the classical shot noise result; in the small bias limit, $\coth\left(\frac{eV}{2k_BT}\right) \approx \frac{2k_BT}{eV}$ and $S_I \approx 4k_BTG$, which represents the thermal fluctuation in the system. Therefore, the excess noise $S_{ex} = 2eVG \coth\left(\frac{eV}{2k_BT}\right) - 4k_BTG$, before accounting for the impedance mismatch as described above.

As explained, it is necessary to calibrate the noise collection efficiency of the measurement setup due to nonidealities in the RF environment. A broadband RF white noise source was employed to test the attenuation of the measurement system integrated over the full bandwidth. It was found that each coaxial cable (and its associated connectors) contributes a 6.5±0.1 dB attenuation. Further, due to the device configuration, the Au leads, wirebonds, and pads introduce additional losses due to the antenna effects and capacitive coupling to ground. Based on these calibration measurements, the bandwidth-integrated RF signal would be further attenuated by 7-8 dB. Therefore, the total signal attenuation beyond that due to the impedance mismatch between the device and the transmission line would be 13.5~14.5 dB. This gives an expected collection efficiency of approximately 3~4%.

To test this, we applied the RF noise measurement technique to two different standards expected to exhibit classical shot noise: Commercially produced Nb-AlO$_x$-Nb tunnel junctions, and a surface-mounted Schottky diode. In the case of the former, assuming the classical shot noise at 10 K with Fano factor of 1 leads to a collection efficiency of 3.34%, as expected. For the Schottky diode at 300 K, again assuming classical shot noise, we find a collection efficiency of 4.77% , not unexpected given that the surface-mount component is lacking the long on-chip leads of the commercial junctions.

We measured the Au-hBN-Au structures at cryostat temperatures from 300 K down to 5 K. We analyzed the data with two approaches. First, after accounting for the impedance mismatch contribution, we fit the measured shot noise to the form $A(2eVG\, coth\left(\frac{eV}{2k_BT}\right) - 4k_BTG)$, so that the noise collection efficiency $A$ can be extracted. In this approach $A$ is the only free parameter, with $T$ assumed to be the cryostat temperature, and $V$ and $G$ measured separately. Fig. 3(c) shows the fitting and residuals at different temperatures for one representative device; results for the other two are quantitatively very similar. The residuals are small compared to the signal throughout the temperature range, showing that the functional form of the measured noise is consistent with expectations of classical shot noise. The inferred efficiency $A$ decays from 3.52% to 2.75% as the cryostat temperature increases from 5K to 300K (Fig. 3(d)). This is quantitatively consistent with the calibration measurements described above. At higher temperatures, the doped substrate underneath the oxide layer is more conductive, and the resulting increased capacitive coupling and consequent change in the RF environment may contribute to enhanced RF signal leakage at higher temperatures. We will test for this effect by fabricating samples on insulating substrates such as sapphire or quartz.

In the second analysis approach, we can infer the effective electronic temperature and measured Johnson-Nyquist noise value based on the shape of the bias-dependent noise intensity itself, which also provides an alternative method to calculate the noise collection efficiency. The excess noise is related to the voltage bias as $S_{ex}/4k_BTG = \frac{eV}{2k_BT}coth\left(\frac{eV}{2k_BT}\right) - 1$, which indicates that the scaled excess noise and scaled voltages at different temperatures all follow the same *xcothx-1* dependence. Following the normalization method in ref [22], the effective electronic temperature $T_e$ and dimensionless noise intensity $S_I^{norm} = S_{ex}/4k_BTG$ can be obtained from the functional form's intersections, as shown in Fig. 4. The electronic temperatures inferred for the device $T_e$ are close but higher than the setting temperatures of the PPMS, as our home-made probe has more thermal lag and greater heatleak from room temperature due to the wiring, compared to a standard PPMS puck. Local resistive heating effects should not be the main reason for the inferred temperature difference, as this would also be expected to distort the functional form of the bias dependence. The normalized extra noise $S_I^{norm}$ shows a consistent *xcothx-1* dependence with the normalized bias $eV/2k_BT$ at different temperatures, which agrees with the shot noise theory prediction very well. The noise collection efficiency based on the extrapolated Johnson-Nyquist noise $4k_BTG$ is consistent with the previous fitting result, as shown in Fig. 3(d). For some devices, there is a weak asymmetry visible in the bias-dependent noise curves, which must originate from asymmetry in the tunneling structure, such as the thin Ti adhesion layer on one side of the junction (which is technically Au/hBN/Ti/Au) or some inevitable chemical residue on one side of the hBN flake from the wet transfer process. A dry transfer method, though more challenging than wet transfer with monolayer hBN, may provide cleaner devices with more symmetric structure and electrical behavior. Devices fabricated without the Ti layer were tested, exhibiting shot noise intensities very close to the case with the Ti adhesion layer, but with better symmetry with bias polarity. For low quality hBN flakes with high defect rates or conductive particles mixed inside, it is possible that the shot noise intensity could deviate from the classical result due to sequential tunneling[23] or vibration-mediated resonant tunneling[24,25]. For clean, intrinsic hBN, we find no obvious deviations from the classical shot noise result.

In conclusion, we fabricated Au/hBN/Au tunnel junctions and studied their shot noise properties. The measured shot noise as a function of bias and temperature is in good agreement with theoretical expectations over a large temperature range, which indicates the potential of such tunnel devices to be applied for thermometry or noise calibration purposes. Further, the clean shot noise result achieved in these normal metal tunnel junctions shows that monolayer hBN can be sufficiently defect-free to function as a promising tunnel barrier for shot noise detection in more complicated systems. Examples of interest for exfoliated tunnel barriers include materials for which other dielectric deposition methods may not be chemically compatible, such as cuprate superconductors and other strongly correlated systems, and spintronic devices where spin accumulation can modify the shot noise.

**ACKNOWLEDGEMENTS**

P.Z., W.J.H., and D.N. gratefully acknowledge support from the US DOE Office of Science/Basic Energy Sciences award DE-FG02-06ER46337. K.W. and T.T. acknowledge support from the Elemental Strategy Initiative conducted by the MEXT, Japan and JSPS KAKENHI Grant Numbers JP26248061, JP15K21722 and JP25106006. The authors thank U. Chandni and P. Gallagher for useful discussion about thin hBN flake identification.

# REFERENCES

[1] M.J.M. de Jong and C.W.J. Beenakker, Phys Rev B **49**, 16070 (1994).

[2] X. Jehl, M. Sanquer, R. Calemczuk, and D. Mailly, Nature **405**, 50 (2000).

[3] R. de-Picciotto, M. Reznikov, M. Heiblum, V. Umansky, G. Bunin, and D. Mahalu, Phys. B Condens. Matter **249–251**, 395 (1998).

[4] L. Saminadayar, D.C. Glattli, Y. Jin, and B. Etienne, Phys Rev Lett **79**, 2526 (1997).

[5] L. Britnell, R.V. Gorbachev, R. Jalil, B.D. Belle, F. Schedin, A. Mishchenko, T. Georgiou, M.I. Katsnelson, L. Eaves, S.V. Morozov, N.M.R. Peres, J. Leist, A.K. Geim, K.S. Novoselov, and L.A. Ponomarenko, Science **335**, 947 (2012).

[6] G.-H. Lee, Y.-J. Yu, X. Cui, N. Petrone, C.-H. Lee, M.S. Choi, D.-Y. Lee, C. Lee, W.J. Yoo, K. Watanabe, T. Taniguchi, C. Nuckolls, P. Kim, and J. Hone, ACS Nano **7**, 7931 (2013).

[7] L. Britnell, R.V. Gorbachev, A.K. Geim, L.A. Ponomarenko, A. Mishchenko, M.T. Greenaway, T.M. Fromhold, K.S. Novoselov, and L. Eaves, Nat. Commun. **4**, 1794 (2013).

[8] A. Mishchenko, J. S. Tu, Y. Cao, R. V. Gorbachev, J. R. Wallbank, M. T. Greenaway, V. E. Morozov, S. V. Morozov, M. J. Zhu, S. L. Wong, F. Withers, C. R. Woods, Y-J. Kim, K. Watanabe, T. Taniguchi, E. E. Vdovin, O. Makarovsky, T. M. Fromhold, V. I. Fal'ko, A. K. Geim, L. Eaves, and K. S. Novoselov, Nat. Nanotechnol. **9**, 808 (2014).

[9] M. Piquemal-Banci, R. Galceran, S. Caneva, M.-B. Martin, R.S. Weatherup, P.R. Kidambi, K. Bouzehouane, S. Xavier, A. Anane, F. Petroff, A. Fert, J. Robertson, S. Hofmann, B. Dlubak, and P. Seneor, Appl. Phys. Lett. **108**, 102404 (2016).

[10] X. Li, W. Cai, J. An, S. Kim, J. Nah, D. Yang, R. Piner, A. Velamakanni, I. Jung, E. Tutuc, S.K. Banerjee, L. Colombo, and R.S. Ruoff, Science **324**, 1312 (2009).

[11] J.W. Suk, A. Kitt, C.W. Magnuson, Y. Hao, S. Ahmed, J. An, A.K. Swan, B.B. Goldberg, and R.S. Ruoff, ACS Nano **5**, 6916 (2011).

[12] L. Wang, I. Meric, P.Y. Huang, Q. Gao, Y. Gao, H. Tran, T. Taniguchi, K. Watanabe, L.M. Campos, D.A. Muller, J. Guo, P. Kim, J. Hone, K.L. Shepard, and C.R. Dean, Science **342**, 614 (2013).

[13] L. Britnell, R.V. Gorbachev, R. Jalil, B.D. Belle, F. Schedin, M.I. Katsnelson, L. Eaves, S.V. Morozov, A.S. Mayorov, N.M.R. Peres, A.H. Castro Neto, J. Leist, A.K. Geim, L.A. Ponomarenko, and K.S. Novoselov, Nano Lett. **12**, 1707 (2012).

[14] U. Chandni, K. Watanabe, T. Taniguchi, and J.P. Eisenstein, Nano Lett. **15**, 7329 (2015).

[15] U. Chandni, K. Watanabe, T. Taniguchi, and J.P. Eisenstein, ArXiv161003189v1 Cond-Matmes-Hall (2016).

[16] S. Jung, M. Park, J. Park, T.-Y. Jeong, H.-J. Kim, K. Watanabe, T. Taniguchi, D.H. Ha, C. Hwang, and Y.-S. Kim, Sci. Rep. **5**, 16642 (2015).

[17] F. Amet, J.R. Williams, A.G.F. Garcia, M. Yankowitz, K. Watanabe, T. Taniguchi, and D. Goldhaber-Gordon, Phys. Rev. B **85**, 073405 (2012).

[18] F. Wu, L. Roschier, T. Tsuneta, M. Paalanen, T. Wang, and P. Hakonen, AIP Conf. Proc. **850**, 1482 (2006).

[19] M. Reznikov, M. Heiblum, H. Shtrikman, and D. Mahalu, Phys. Rev. Lett. **75**, 3340 (1995).

[20] P.J. Wheeler, J.N. Russom, K. Evans, N.S. King, and D. Natelson, Nano Lett. **10**, 1287 (2010).

[21] D. Rogovin and D.. Scalapino, Ann. Phys. **86**, 1 (1974).

[22] L. Spietz, K.W. Lehnert, I. Siddiqi, and R.J. Schoelkopf, Science **300**, 1929 (2003).

[23] L.Y. Chen and C.S. Ting, Phys. Rev. B **46**, 4714 (1992).

[24] M. Kumar, R. Avriller, A.L. Yeyati, and J.M. van Ruitenbeek, Phys. Rev. Lett. **108**, 146602 (2012).

[25] D. Harbusch, D. Taubert, H.P. Tranitz, W. Wegscheider, and S. Ludwig, Phys. Rev. Lett. **104**, 196801 (2010).

**FIGURE LEGENDS**

FIG. 1. Optical images of devices. (a) Optical micrograph of an hBN flake (part of which is a monolayer) on 300 nm silicon oxide coated wafer. (b) The same hBN flake after transfer onto bottom Au electrodes (outlined in red for clarity). The image is taken with a green filter to improve the optical contrast. (c) Finished junction device after deposition of top Au electrodes.

FIG. 2. Electrical characterizations of devices. (a) I-V curves of three different devices acquired at T = 5 K. (b) Corresponding dI/dV curves of the devices in (a) at the same temperature.

FIG. 3. Shot noise measurement on Au/hBN/Au junctions. (a) Schematic circuit of the shot noise measurement setup. (b) Effective electrical circuit of the RF signal transmission. $R$ is the sample's resistance; $Z_0$ is transmission line/amplifier's load resistance; $C_1$ is the capacitance of the tunnel junction; $C_2$ is additional capacitance of the on-chip Au leads and bonding pads, and $L$ is the effective inductance of Au leads. (c) Fits (solid lines) to the measured shot noise intensity (open symbols) and below, the corresponding residuals at different temperatures. The noise collection efficiency A is the only adjustable parameter in the fit. (d) Extracted noise collection efficiencies at different temperatures. The black dots are the noise collection efficiencies obtained from the fitting. The red dots are the noise collection efficiencies extracted from the intersection analysis of Fig. 4, as described in the main text.

FIG. 4. Effective temperatures and normalized shot noise of junction sample. (a) Effective temperatures of sample versus PPMS setting temperatures. Inset: theoretical plot of classical shot noise versus bias, where the effective temperature and noise normalization factor $4k_BTG$ can be extracted from the intersection. (b) Normalized shot noise versus bias at different temperatures. The dashed line indicates the theoretical prediction. (The applied bias range is not large to apply this analysis procedure to the 300 K data.)

**FIGURE 1**

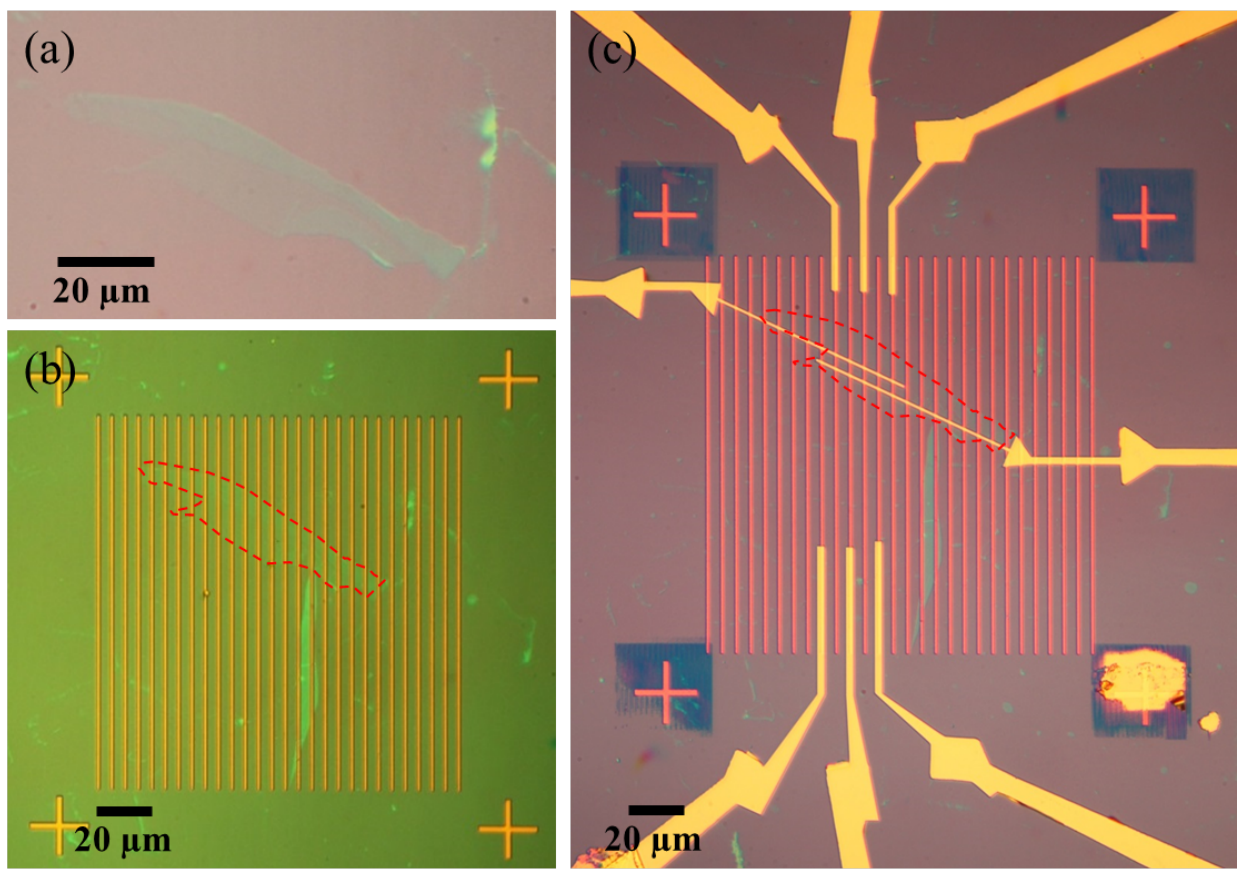

**FIGURE 2**

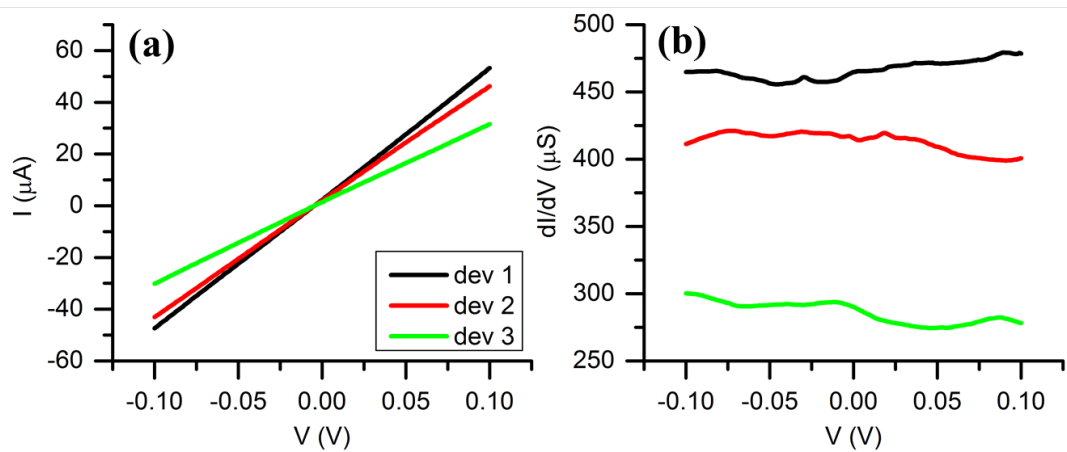

[Insert Running title of <72 characters]

**FIGURE 3**

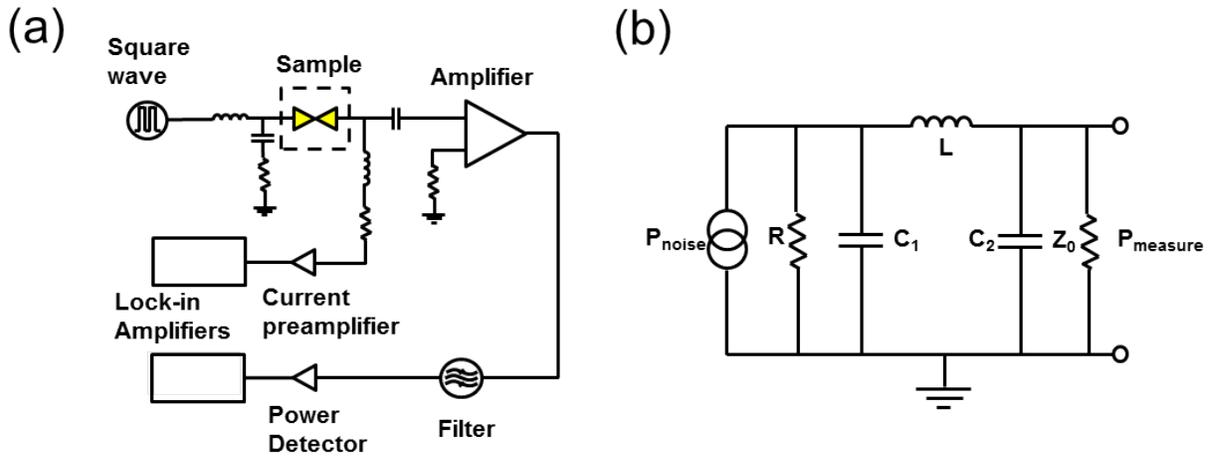
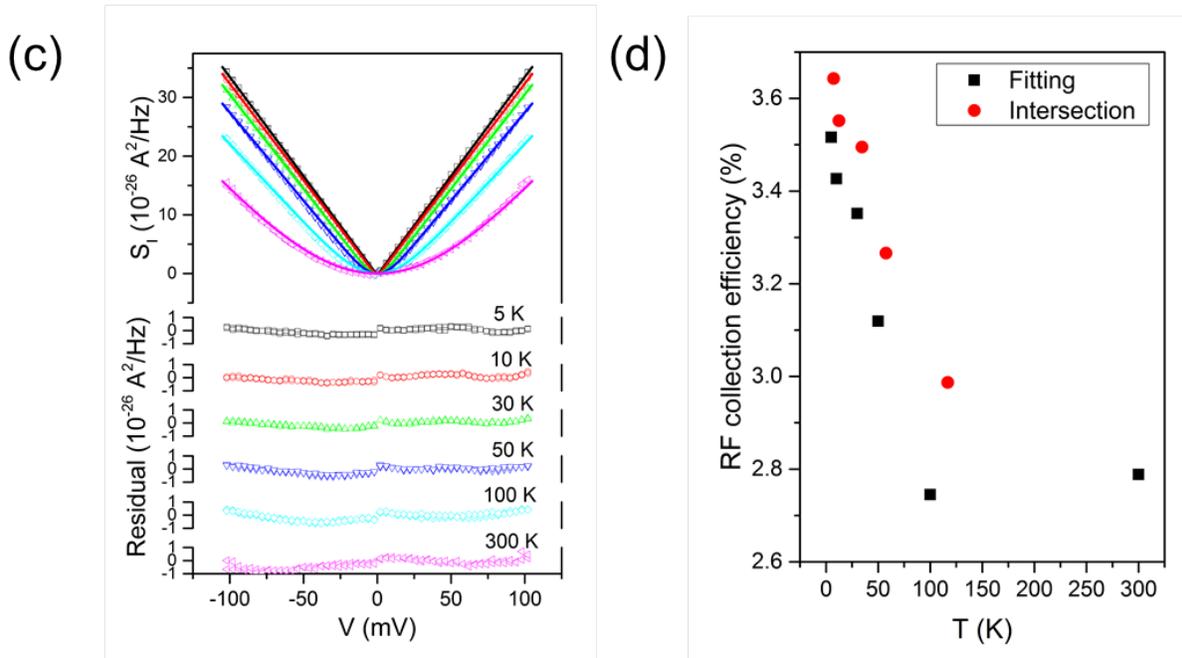

[Insert Running title of <72 characters]

**FIGURE 4**

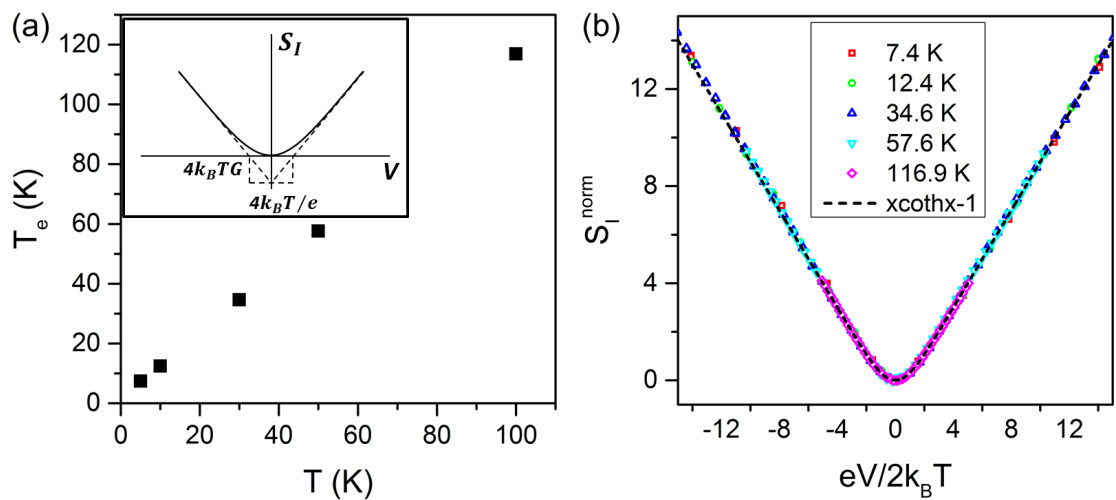